\documentclass[conference]{IEEEtran}

\usepackage[T1]{fontenc} %
\usepackage[utf8]{inputenc} %
\usepackage[english]{babel} %
\usepackage[
    backend=biber,
    style = ieee,
    citestyle = ieee-comp,
    sortlocale=en_US,
    sortcites=true,
    url=true,
    eprint=true,
    giveninits=false,
    dashed=false,
    minnames=1,
    maxnames=5
]{biblatex}
\usepackage{amsmath,amssymb,amsfonts}
\usepackage{graphicx}
\usepackage{pdfpages}
\usepackage{textcomp}
\usepackage{xcolor}
\usepackage{xspace} %
\usepackage[inline]{enumitem} %
\usepackage{amsthm,thmtools} %
\usepackage{tabularx} %
\usepackage{colortbl}
\usepackage{booktabs} %
\usepackage{siunitx} %
\usepackage{csquotes} %
\usepackage[noEnd]{algpseudocodex} %
\makeatletter
\let\MYcaption\@makecaption
\makeatother
\usepackage{subcaption} %
\makeatletter
\let\@makecaption\MYcaption
\makeatother
\usepackage[hidelinks]{hyperref} %
\usepackage[capitalize,nameinlink]{cleveref} %
\AtEveryBibitem{%
    \clearname{editor}%
    \clearfield{series}%
    \clearfield{isbn}%
    \clearfield{issn}%
    \clearfield{day}%
    \clearfield{month}%
    \clearfield{place}%
    \clearlist{location}%
    \clearfield{pages}%
    \clearfield{volume}%
    \clearfield{number}%
    \clearfield{eprintclass}%
    \ifentrytype{software}{%
    }{%
        \clearfield{url}%
        \clearfield{urlyear}%
        \clearfield{urldate}%
    }%
}
\DeclareSourcemap{
    \maps[datatype=bibtex]{
        \map{
            \step[fieldsource=doi, match=\regexp{.+/arXiv\..+}, final]
            \step[fieldsource=eprint, match=\regexp{.+}, final]
            \step[fieldset=doi, null]
        }
        \map{
            \step[fieldsource=doi, match=\regexp{.+}, final]
            \step[fieldset=eprint, null]
            \step[fieldset=archiveprefix, null]
            \step[fieldset=eprinttype, null]
        }
    }
}
\graphicspath{ {figures/} }
\setlist[enumerate]{label=(\arabic*)} %
\declaretheoremstyle[
    bodyfont=\itshape
]{example-style}
\declaretheorem[
    name=Example,
    style=example-style,
    numbered=unless unique,
]{example}
\makeatletter
\long\def\@makecaption#1#2{%
\ifx\@captype\@IEEEtablestring%
\footnotesize\bgroup\par\centering\@IEEEtabletopskipstrut{\normalfont\footnotesize {#1.}\nobreakspace\scshape #2}\par\addvspace{0.5\baselineskip}\egroup%
\@IEEEtablecaptionsepspace
\else
\@IEEEfigurecaptionsepspace
\setbox\@tempboxa\hbox{\normalfont\footnotesize {#1.}\nobreakspace #2}%
\ifdim \wd\@tempboxa >\hsize%
\setbox\@tempboxa\hbox{\normalfont\footnotesize {#1.}\nobreakspace}%
\parbox[t]{\hsize}{\normalfont\footnotesize \noindent\unhbox\@tempboxa#2}%
\else%
\hbox to\hsize{\normalfont\footnotesize\hfil\box\@tempboxa\hfil}%
\fi\fi}
\makeatother
\crefname{section}{Sec.}{Sec.}
\Crefname{section}{Sec.}{Sec.}
\crefname{example}{Ex.}{Ex.}
\Crefname{example}{Ex.}{Ex.}
\newcolumntype{R}{>{\raggedleft\arraybackslash}X}
\newcolumntype{C}{>{\centering\arraybackslash}X}

\definecolor{TUM_blue}{RGB}{0,101,189}
\colorlet{TUM_black}{black}
\colorlet{TUM_white}{white}
\definecolor{TUM_darkblue}{RGB}{0,82,147}
\colorlet{TUM_darkblue100}{TUM_darkblue}
\colorlet{TUM_darkblue80}{TUM_darkblue100!80}
\colorlet{TUM_darkblue50}{TUM_darkblue100!50}
\colorlet{TUM_darkblue20}{TUM_darkblue100!20}
\definecolor{TUM_verydarkblue}{RGB}{0,51,89}
\colorlet{TUM_verydarkblue100}{TUM_verydarkblue}
\colorlet{TUM_verydarkblue80}{TUM_verydarkblue100!80}
\colorlet{TUM_verydarkblue50}{TUM_verydarkblue100!50}
\colorlet{TUM_verydarkblue20}{TUM_verydarkblue100!20}
\colorlet{TUM_darkgrey}{TUM_black!80}
\colorlet{TUM_grey}{TUM_black!50}
\colorlet{TUM_lightgrey}{TUM_black!20}
\definecolor{TUM_beige}{RGB}{218,215,203}
\definecolor{TUM_orange}{RGB}{227,114,34}
\definecolor{TUM_green}{RGB}{162,173,0}
\definecolor{TUM_verylightblue}{RGB}{152,198,234}
\definecolor{TUM_lightblue}{RGB}{100,160,200}

\title{Guiding Compiler Optimizations for Neutral Atom Quantum Computers Through Visualizations}

\author{
    \IEEEauthorblockN{
        Yannick Stade\IEEEauthorrefmark{1},
        and Robert Wille\IEEEauthorrefmark{1}\IEEEauthorrefmark{2}
    }
    \IEEEauthorblockA{\IEEEauthorrefmark{1}%
    Chair for Design Automation, Technical University of Munich, Munich, Germany
    }
    \IEEEauthorblockA{\IEEEauthorrefmark{2}%
    MQSC, Garching near Munich, Germany
    }
    yannick.stade@tum.de, %
    robert.wille@tum.de\\
    \href{https://www.cda.cit.tum.de/research/quantum}{www.cda.cit.tum.de/research/quantum}
}
\hypersetup{ %
    pdftitle={Guiding Compiler Optimizations for Neutral Atom Quantum Computers Through Visualizations},
    pdfsubject={IEEE International Conference on Quantum COmputing and Engineering (QCE) 2026},
    pdfauthor={
        Yannick Stade,
        Robert Wille
    }
}

\begin{document}

\maketitle

\begin{abstract}
    The scale of Neutral Atom (NA) quantum computers requires automated compilation tools.
    Designing the required heuristic methods demands a deep understanding of complex hardware trade-offs, for which visualizations can provide crucial insights.
    This work introduces \emph{NAViz}, the first publicly available app to visualize quantum computations on NA devices in real-time.
    A case study demonstrates how \emph{NAViz} was instrumental in identifying and resolving inefficiencies in an existing compilation strategy, leading to a new, more performant one.
    The tool is available as part of the \emph{Munich Quantum Toolkit}~(MQT) at \url{https://github.com/munich-quantum-toolkit/naviz}.
\end{abstract}


\section{Introduction \& Background}\label{sec:introduction}

Neutral Atom (NA) quantum computers offer scalability and flexible connectivity by rearranging qubits with optical tweezers~\cite{bluvsteinArchitecturalMechanismsUniversal2025}.
This makes \emph{placement} (assigning atoms to sites) and \emph{routing} (moving atoms) crucial compilation tasks~\mbox{\cite{patelGRAPHINEEnhancedNeutral2023,schmidHybridCircuitMapping2024,silverQomposeTechniqueSelect2024,wangQPilotFieldProgrammable2024,ludmirPARALLAXCompilerNeutral2024,tanCompilationDynamicallyFieldProgrammable2025,stadeRoutingAwarePlacementZoned2025,constantinidesOptimalRoutingProtocols2024,tanCompilingQuantumCircuits2024,stadeOptimalStatePreparation2024}}.

The scale of current devices requires heuristic methods~\cite{stadeRoutingAwarePlacementZoned2025,linReuseAwareCompilationZoned2024}.
Designing these heuristics requires deep insights, for which visualizations have proven successful in other domains~\cite{willeRevVisVisualizationStructures2014,hansenSemanticZoomMiniMaps2025}.

This work, for the first time, showcases how visualizations help to better understand and improve compilation strategies for NA devices.
To this end, we introduce \emph{NAViz}, a visualization app designed to analyze various compilation strategies and provide intuition on their effects.
We then apply NAViz to a routing problem where an intuitively superior strategy paradoxically results in worse performance.
NAViz reveals the cause of this inefficiency, which ultimately led to the development of a novel, better-performing strategy.

\section{Visualization app}\label{sec:tool}

An NA quantum computation consists of a sequence of quantum gates interleaved with atom rearrangements.
Minimizing rearrangement time is critical~\cite{schmidComputationalCapabilitiesCompiler2024,stadeRoutingAwarePlacementZoned2025}.
Visualizing atom movements can reveal inefficiencies in routing strategies.
To this end, we developed \emph{NAViz}, a visualization app for quantum computations on NA devices, implemented in Rust to facilitate \mbox{real-time} rendering.
NAViz takes the quantum computation and device architecture as input, with customizable appearance settings.
It provides an interactive real-time visualization with adjustable speed, a timeline for detailed analysis, and an option to export animations as video files.

\begin{figure}[t]
    \centering
    \includegraphics[width=.7\linewidth,trim=0 72pt 0 36pt]{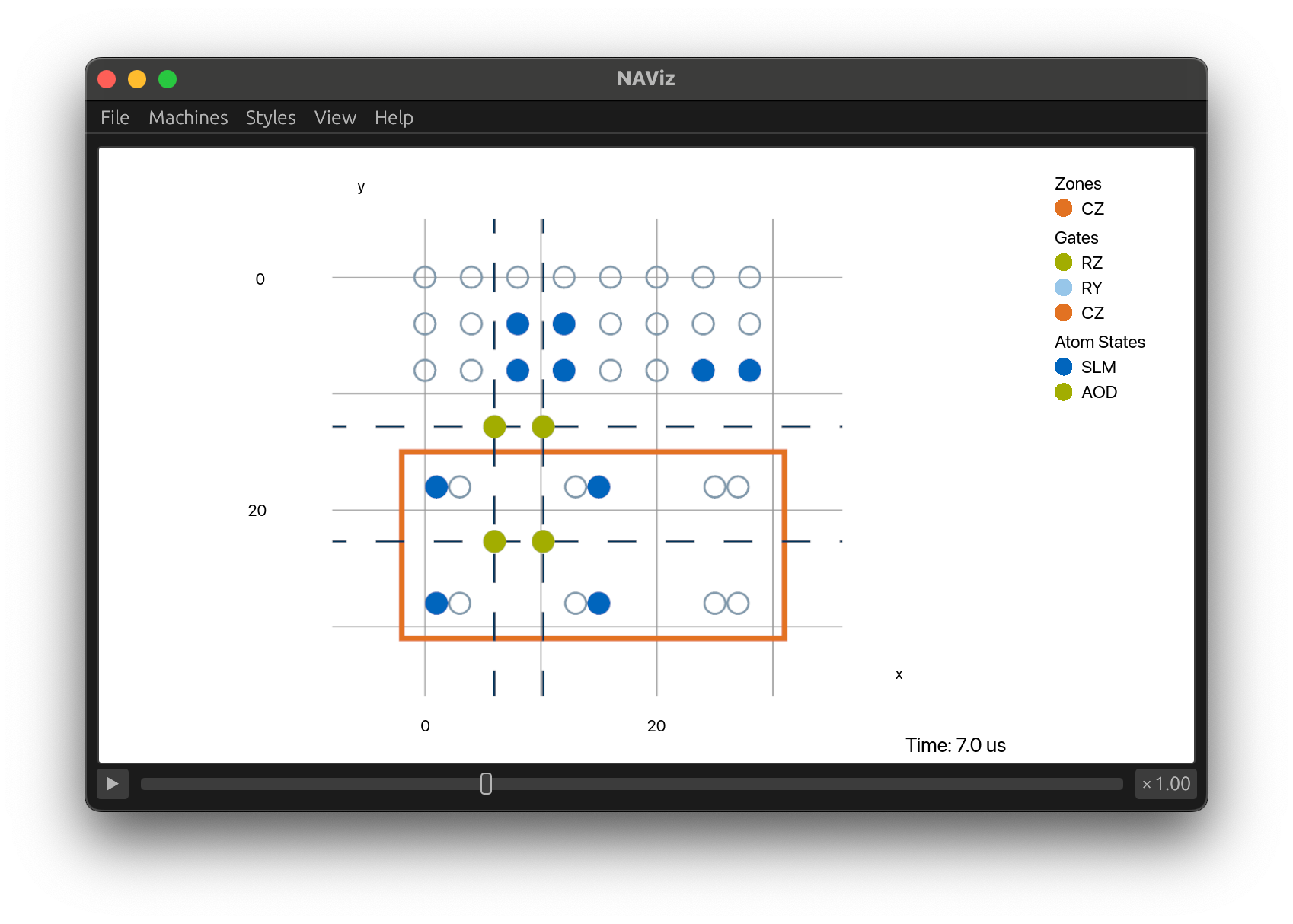}\\\vspace{-6pt}%
    \caption{Screenshot of the visualization app NAViz while animating a quantum computation on an NA device.}
    \label{fig:naviz}
\end{figure}

\begin{example}\label{exp:naviz}
    \Cref{fig:naviz} shows a screenshot of NAViz while animating a quantum computation.
    In the background, gray circles depict the potential static locations for atoms on the device.
    The atoms themselves are shown as colored circles: blue for static atoms and green for those currently being rearranged.
    The timeline slider at the bottom allows the user to navigate through the time.
\end{example}

\section{Compilation Aspect: Relaxed Routing}\label{sec:relaxed routing}

The rearrangement of atoms on NA devices is facilitated by static \emph{Spatial Light Modulator}~(SLM) traps and two orthogonal \emph{Acusto-Optic Deflectors}~(AODs).
These create an adjustable 2D grid of traps, where each column and row can be controlled independently~\cite{bluvsteinQuantumProcessorBased2022}.
To rearrange an atom, its corresponding column and row are activated picking it up, shifted to the new position, and then deactivated to drop it off.
While multiple atoms can be rearranged in parallel, columns (rows) are activated/deactivated as a whole, creating complex \emph{rearrangement constraints} that can lead to serializing rearrangements~\cite{stadeAbstractModelEfficient2024}.

Furthermore, since during a rearrangement, columns (rows) cannot cross, the relative order of the atoms in a rearrangement cannot change.
More precisely, any two atoms \(a_1\) and \(a_2\) can only be rearranged in parallel if they satisfy the conditions:
\begin{align}
    x^{(\mathrm{old})}_{a_1} = x^{(\mathrm{old})}_{a_2} &\Longleftrightarrow x^{(\mathrm{new})}_{a_1} = x^{(\mathrm{new})}_{a_2}\label{eq:x same}\\
    y^{(\mathrm{old})}_{a_1} = y^{(\mathrm{old})}_{a_2} &\Longleftrightarrow y^{(\mathrm{new})}_{a_1} = y^{(\mathrm{new})}_{a_2}\label{eq:y same}\\
    x^{(\mathrm{old})}_{a_1} < x^{(\mathrm{old})}_{a_2} &\Longleftrightarrow x^{(\mathrm{new})}_{a_1} < x^{(\mathrm{new})}_{a_2}\label{eq:x order}\\
    y^{(\mathrm{old})}_{a_1} < y^{(\mathrm{old})}_{a_2} &\Longleftrightarrow y^{(\mathrm{new})}_{a_1} < y^{(\mathrm{new})}_{a_2}\label{eq:y order}
\end{align}

However, these \emph{strict} conditions can be slightly \emph{relaxed} by exploiting the fact that, when the first atom is picked up, it can already be moved past other atoms.

This flexibility allows dropping conditions \cref{eq:x order,eq:y order}.
However, as shown in \cref{tab:results}, this \emph{relaxed} strategy paradoxically increased rearrangement times compared to the \emph{strict} one.

\section{Learning from the Visualization}\label{sec:insights}

\begin{figure}[t]
    \centering
    \includegraphics[width=.8\linewidth]{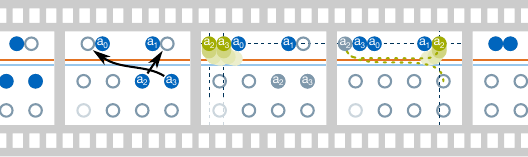}\\
    \raggedright\vspace{-40pt}\hspace{55pt}%
    \begin{subfigure}[t]{51pt}
            \caption{\raggedright}
            \label{subfig:objective}
    \end{subfigure}%
    \begin{subfigure}[t]{51pt}
            \caption{\raggedright}
            \label{subfig:first atom}
    \end{subfigure}%
    \begin{subfigure}[t]{20pt}
            \caption{\raggedright}
            \label{subfig:second atom}
    \end{subfigure}\\\vspace{20pt}%
    \caption{
        The black arrows in \cref{subfig:objective} indicate the desired rearrangement of two atoms.
        \Cref{subfig:first atom} depicts the first movement and \cref{subfig:second atom} the second.
        The dashed green line in \cref{subfig:second atom} illustrates the distance traveled by atom \(a_2\).
    }
    \label{fig:relaxed routing}
\end{figure}

To shed light onto these counter-intuitive results, we employed NAViz to visualize various rearrangements produced by the relaxed strategy.
This analysis revealed inefficient detours.

\begin{example}\label{exp:hybrid strategy}
    \Cref{fig:relaxed routing} depicts a scenario of relaxed routing discovered through NAViz.
    Here, both atoms are picked up simultaneously.
    Atom~\(a_3\) is moved to its target position.
    Meanwhile atom~\(a_2\) must
    \begin{enumerate*}
        \item stay in the same row (cf.~\cref{eq:y same}), and
        \item remain left of atom~\(a_3\) (cf.~\cref{eq:x order})
    \end{enumerate*}%
    ---effectively causing a long detour for atom~\(a_2\) (dashed green line in \cref{subfig:second atom}).
\end{example}

Thanks to NAViz, a better \emph{hybrid} strategy combining the strict and relaxed approaches could be derived to mitigate those inefficient detours.
The strategy first computes a strict routing and then greedily merges rearrangement steps if doing so reduces the total overhead under relaxed conditions.
Evaluations on a set of benchmarks~\cite{quetschlichMQTBenchBenchmarking2023} demonstrate that this hybrid strategy generally outperforms both others, as summarized in \cref{tab:results}.

\begin{table}[t]
    \setlength{\aboverulesep}{0.4pt} %
    \setlength{\belowrulesep}{0.4pt} %
    \caption{Results of Different Routing Strategies}\vspace{-8pt}
    \label{tab:results}
    \small
    \begin{tabular}{%
        p{49pt}%
        S[table-format=4.0]%
        |S[table-format=4.1]%
        |c
        S[table-format=4.1]%
        |c
        S[table-format=4.1]%
    }
            \toprule
            \multicolumn{2}{l|}{\textbf{Benchmark}} & {\textbf{Strict}} & \multicolumn{2}{c|}{\textbf{Relaxed}} & \multicolumn{2}{c}{\textbf{Hybrid}} \\[-1pt]
            \multicolumn{2}{r|}{Num.} & {Rearr.} & \multicolumn{2}{c|}{Rearr.} & \multicolumn{2}{c}{Rearr.} \\[-1pt]
            \multicolumn{2}{r|}{Qubits} & {T. [\si{\milli\second}]} & \multicolumn{2}{c|}{Time [\si{\milli\second}]} & \multicolumn{2}{c}{Time [\si{\milli\second}]} \\
            \midrule
            graphstate &  100 &   22.4 & \cellcolor{TUM_orange}$\nearrow$ & \cellcolor{TUM_orange}   26.2 & \cellcolor{TUM_green}$\searrow$  & \cellcolor{TUM_green}   22.1 \\
            &  200 &   53.0 & \cellcolor{TUM_orange}$\nearrow$ & \cellcolor{TUM_orange}   67.5 & \cellcolor{TUM_green}$\searrow$  & \cellcolor{TUM_green}   51.8 \\
            &  500 &  133.8 & \cellcolor{TUM_orange}$\nearrow$ & \cellcolor{TUM_orange}  198.5 & \cellcolor{TUM_green}$\searrow$  & \cellcolor{TUM_green}  132.8 \\
            & 1000 &  320.5 & \cellcolor{TUM_orange}$\nearrow$ & \cellcolor{TUM_orange}  517.4 & \cellcolor{TUM_orange}$\nearrow$ & \cellcolor{TUM_orange} 320.9 \\
            & 2000 &  802.7 & \cellcolor{TUM_orange}$\nearrow$ & \cellcolor{TUM_orange} 1322.9 & \cellcolor{TUM_green}$\searrow$  & \cellcolor{TUM_green}  802.1 \\
            & 5000 & 2588.8 & \cellcolor{TUM_orange}$\nearrow$ & \cellcolor{TUM_orange} 4147.7 & \cellcolor{TUM_green}$\searrow$  & \cellcolor{TUM_green} 2582.3 \\
            qft &  500 & 2324.5 & \cellcolor{TUM_orange}$\nearrow$ & \cellcolor{TUM_orange} 2354.5 & \cellcolor{TUM_green}$\searrow$  & \cellcolor{TUM_green} 2287.1 \\
            & 1000 & 4937.9 & \cellcolor{TUM_orange}$\nearrow$ & \cellcolor{TUM_orange} 4986.7 & \cellcolor{TUM_green}$\searrow$  & \cellcolor{TUM_green} 4855.8 \\
            vqe two local &   50 &  631.9 & \cellcolor{TUM_orange}$\nearrow$ & \cellcolor{TUM_orange}  670.4 & \cellcolor{TUM_green}$\searrow$  & \cellcolor{TUM_green}  624.7 \\
            &  100 & 2423.1 & \cellcolor{TUM_orange}$\nearrow$ & \cellcolor{TUM_orange} 2688.8 & \cellcolor{TUM_green}$\searrow$  & \cellcolor{TUM_green} 2394.6 \\
            &  200 & 8477.0 & \cellcolor{TUM_orange}$\nearrow$ & \cellcolor{TUM_orange} 9757.0 & \cellcolor{TUM_green}$\searrow$  & \cellcolor{TUM_green} 8349.7 \\
            wstate &  500 &  410.8 & \cellcolor{TUM_green}$\searrow$  & \cellcolor{TUM_green}   406.3 & \cellcolor{TUM_green}$\searrow$  & \cellcolor{TUM_green}  410.5 \\
            & 1000 &  777.7 & \cellcolor{TUM_green}$\searrow$  & \cellcolor{TUM_green}   772.6 & \cellcolor{TUM_green}$\searrow$  & \cellcolor{TUM_green}  777.1 \\
            \midrule
            \(\varnothing\) &    & 1838.8 & \cellcolor{TUM_orange}$\nearrow$ & \cellcolor{TUM_orange} 2147.4 & \cellcolor{TUM_green}$\searrow$  & \cellcolor{TUM_green} 1816.3 \\
            \bottomrule
    \end{tabular}
\end{table}

\section{Conclusions}\label{sec:conclusions}

We introduced \emph{NAViz}, an open-source visualization tool for NA quantum computers.
A case study demonstrated its utility in identifying compiler inefficiencies and developing a superior, hybrid routing strategy.
The tool is available in open-source as part of the MQT under \url{https://github.com/munich-quantum-toolkit/qmap/}.

\subsection*{Acknowledgments}\label{sec:ack}
\footnotesize
During the preparation of the code and manuscript, the authors used GitHub Copilot, powered by Claude's Sonnet 4.6 to improve code, spelling, grammar, clarity, and readability.
Afterward, the authors reviewed and edited the content as needed.
The authors take full responsibility for the final content.

The project leading to this publication has received funding from the European Research Council (ERC) under the European Union’s Horizon 2020 research and innovation program (grant agreement No. 101001318), the Munich Quantum Valley, which is supported by the Bavarian state government with funds from the Hightech Agenda Bayern Plus, the Deutsche Forschungsgemeinschaft (DFG, German Research Foundation, 563436708), and the BMFTR under grant number 01MQ25001I (FullStaQD).

\renewcommand*{\bibfont}{\scriptsize} %
\printbibliography

\end{document}